\documentclass[%
reprint,
 amsmath,amssymb,
 aps,
]{revtex4-1}

\usepackage{graphicx}
\usepackage{dcolumn}
\usepackage{bm}
\usepackage{color}
\usepackage{multirow}
\usepackage{float}

\begin{document}

\title{Effect of Frustrated Exchange Interactions and Spin-half Impurity on the Electronic Structure of Strongly Correlated NiFe$_2$O$_4$}

\author{Kodam Ugendar$^{1, \S}$, S. Samanta$^{2, \S}$, Sudhendra Rayaprol$^3$, V. Siruguri$^3$, G. Markandeyulu$^{1}$, B. R. K. Nanda$^{2,\star}$}
\affiliation{$^1$ Advanced Magnetic Materials Laboratory, Department of Physics, Indian Institute of Technology Madras, Chennai – 600 036, India}
\affiliation{$^2$Condensed Matter Theory and Computational Lab, Department of Physics, Indian Institute of Technology Madras, Chennai – 600 036, India}
\email{nandab@iitm.ac.in}
\email{$^\S$ equal contribution}
\affiliation{$^3$ UGC-DAE Consortium for Scientific Research, Mumbai center, Bhabha Atomic Research Centre, Trombay, Mumbai - 400 085, India}

\begin{abstract}
Spin-polarized density functional calculations, magnetization, and neutron diffraction (ND) measurements are carried out to investigate the magnetic exchange interactions and strong correlation effects in Yb substituted inverse spinel nickel ferrite. In the pristine form, the compound is found to be a mixed insulator under the Zaanen-Sawatzky-Allen classification scheme as it features both charge transfer and Mott insulator mechanism.  Estimation of magnetic exchange couplings reveals that both  octahedral-octahedral and octahedral-tetrahedral spin-spin interactions are antiferromagnetic.
This is typical of spin-frustrated triangular lattice with one of the vertices occupied by tetrahedral spins and remaining two are occupied by octahedral spins. However, since the octahedral-tetrahedral interaction is dominant, it leads to a forced
 parallel alignment of the spins at the octahedral site
which is in agreement with the results of ND measurements. The substituent Yb is found to be settled in +3 charge state, as confirmed from the XPS measurements, to behave like a spin-half impurity carried by the localized  $f_{z(x^2-y^2)}$ orbital. The impurity $f$ spin  significantly weakens the antiferromagnetic coupling with the spins at the tetrahedral site,  which explains the experimental observation of fall in Curie temperature with Yb substitution.
\end{abstract}

\maketitle

\section{\label{sec:Intro}Introduction}
The cubic inverse spinel NiFe$_2$O$_4$ (NFO) has been extensively investigated in the context of nanomagnetism \cite{Wang4}, spin-filtering \cite{Matzeni,Nuala}, spintronics \cite{Luders} and multiferroics \cite{Verma}. In addition, it exhibits unusual electronic and magnetic properties when Fe$^{3+}$/Ni$^{2+}$ ions 
at the octahedral sites
are partially substituted by other transition metal (M) ions, rare earth (R) ions or ions of non-transition elements \cite{Chappert, Kamala, KamalaJPC, RezlescuJPCM, RezlescuSSU, RezlescuPSS}. Collinear N$\acute{e}$el type ferrimagnetic structure of NFO yields to triangular Yafet-Kittel structure upon substantial Cr substitution at the Fe cations at the octahedral sites \cite{Chappert}. The octahedra containing Fe$^{3+}$ ions in NFO, when partially substituted by rare-earth (R$^{3+}$), become non-centrosymmetric to make the compound ferroelectric. Experimentally it has been shown that, substituents like Sm$^{3+}$ and Ho$^{3+}$  induce magnetoelectric effect in NFO \cite{KamalaJPC}. 
\vspace{6pt}
\\
Significant changes in the electronic, magnetic and structural behavior of Ni-Zn ferrite upon diluting with several rare earth ions have been observed \cite{RezlescuJPCM, RezlescuSSU, RezlescuPSS}.  With substitution of 2\% of Fe by R (= Yb, Er, Dy, Tb, Gd, Sm and Ce) in  Ni$_{0.7}$Zn$_{0.3}$Fe$_{2}$O$_4$, while lattice has been reported to expand and resistivity has increased, both magnetization and Curie temperature (T$_C$) have decreased \cite{RezlescuJPCM, RezlescuSSU, RezlescuPSS}. Larger ionic radii of R$^{3+}$ ions cause lattice expansion and the 4\textit{f} elections are more localized than the itinerant 3\textit{d} electrons and hence, the resistivity increases \cite{RezlescuJPCM, RezlescuSSU, RezlescuPSS}. 
The reported value of T$_C$ of NFO is 853 K \cite{Chikazumi, Kamala}.
A decrease in T$_C$ upon the partial substitution of R$^{3+}$ for Fe$^{3+}$ in
NFO has been reported from our lab \cite{Kamala, KamalaJPC}.
In Ni$_{2}$Fe$_{1.925}$R$_{0.075}$O$_4$ compounds, the T$_C$ decreases to 775 K,  812 K and 839 K respectively for Dy$^{3+}$ \cite{Kamala}, Ho$^{3+}$ and Sm$^{3+}$ substitutions \cite{KamalaJPC}.
However, there are no concrete mechanisms and evidences to explain the decrease in magnetization and Curie temperature, even though qualitatively it has been attributed to weaker R-Fe exchange coupling replacing the stronger Fe-Fe exchange coupling \cite{Chikazumi,Smith}.
\vspace{6pt}
\\
In this paper, results from density functional theory (DFT) calculations and  experimental studies  are presented and analyzed to explain the electronic and magnetic structures of Yb substituted NFO \textit{viz.} NiFe$_{2-x}$Yb$_{x}$O$_{4}$ ($x$ = 0, 0.05, 0.075). The reasons for choosing Yb were manifold: (a) Structural distortion is expected to be weak or negligible, since the radius of Yb$^{3+}$ ion (0.86 \AA) is smaller compared to those of the other rare earth ions. (b) Yb ion can stabilize in +2 and +3 charge states. (c) Yb$^{3+}$ is magnetic and has lower spin moment compared to the other R$^{3+}$ (R = Gd, Tb, Dy, Ho, Er, Tb) ions \cite{Chikazumi} and hence, large reduction in magnetization as well as Curie temperature. (d) Yb$^{3+}$ is expected to provide a spin-half \textit{f} impurity state.  Therefore, it serves as a model system to study host (\textit{d} spin)-impurity (\textit{f} spin) magnetic interactions.
\vspace{6pt}
\\
Experimentally, X-ray photoelectron spectroscopy (XPS), Raman spectroscopy and ND measurements are performed and theoretically, spin-polarized band structure is calculated to explain the electronic  structure of NiFe$_{2-x}$Yb$_{x}$O$_{4}$. In addition, various magnetic exchange couplings are estimated from the total energies of several possible magnetic configurations so that the spin-spin interactions in this compound can be better understood. Emphasis is given on the magnetic coupling of Yb and Fe spins and its effect on the net magnetization as well as T$_C$ of this inverse spinel compound.
\vspace{6pt}
\\
The compound NFO behaves as a mixed insulator under the Zaanen-Sawatzky-Allen classification scheme with the trait of both Mott and charge transfer insulating phenomena.  It is found that the spins of Fe$^{3+}$ and Yb$^{3+}$ ions at the octahedral sites prefer to align antiparallel. However, stronger antiferromagnetic coupling with the spins at the tetrahedral sites forces them to align parallel in order to avoid spin frustration and thereby stabilizing ferrimagnetism in NFO. The +3 charge state, confirmed from XPS measurements and DFT studies, makes Yb a spin-half ion with the $f_{z(x^2-y^2)}$ orbital carrying the unpaired spin. Experimental observation of the decrease of Curie temperature by approximately 14 K, with 7.5 \% Yb substitution is attributed to the fact that the Yb spin significantly weakens the antiferromagnetic coupling with the neighboring spins at the tetrahedral site and marginal enhancement of the same with the spins at the neighboring octahedral sites.  
\vspace{6pt}
\\
The rest of the paper is organized as follows.  Section~\ref{sec:ExpStds} and ~\ref{sec:DFT-Stds} report experimental and theoretical investigations respectively. Section ~\ref{sec:Instrument} presents the experimental details involving material synthesis (NiFe$_{2-x}$Yb$_{x}$O$_{4}$ ($x$ = 0, 0.05, 0.075)) and characterization. Section~\ref{sec:ExpStruct} provides the basic crystal structure information of the Yb substituted NFO based on X-ray diffraction (XRD) and Raman spectroscopy studies. Section ~\ref{sec:ExpXPSND} analyzes the results from XPS and ND studies. Computational details are reported in section~\ref{sec:Comp-dtls}, and corresponding results are presented in Section~\ref{sec:ElcMag-dtls}. In section ~\ref{sec:MagExch}, the magnetic exchange interactions are presented to explain the effect of Yb-\textit{f} state on host NFO. Finally, the results are summarized in section~\ref{sec:Summary}. 
\section{\label{sec:ExpStds}Experimental studies}

\subsection{\label{sec:Instrument}Synthesis and Characterization of the compounds}
Polycrystalline samples of NiFe$_{2-x}$Yb$_{x}$O$_{4}$ ($x$ = 0, 0.05, 0.075) were prepared starting from NiO (99.96\% pure), Fe$_{2}$O$_{3}$ and Yb$_{2}$O$_{3}$ (99.99\% pure), by solid state reaction method.  The powders of the starting materials were ground in an agate mortar and pestle for 3 h and heat treated in air at 1200 $^{0}$C for 12 h.  The phase formation of each of the samples was confirmed by taking powder XRD patterns employing a PANalytical (X’pert PRO) x-ray diffractometer with Cu K$_{\alpha}$ radiation.  Raman active vibrational modes in the samples were recorded using a Horiba Jobin Yvon HR800 UV: Raman Division, Raman spectrometer, with an excitation wave length of 633 nm, in the wave number range 180 to 750 cm$^{-1}$.  XPS and ND measurements were carried out on the compound with the highest concentration of Yb, NiFe$_{1.925}$Yb$_{0.075}$O$_{4}$. The Fe 2\textit{p}, Ni 2\textit{p}, Yb 4\textit{d} and O 1\textit{s} XPS spectra were recorded with micro focused monochromatic X-ray source (Design: Sigma Probe) having an energy resolution of 0.47 eV at FWHM.  Spectroscopic studies were carried out on thin pellets;  the binding energies were charge corrected with reference to C 1\textit{s} energy level at 284.5 eV.  Magnetization data were obtained by employing a vibrating sample magnetometer (VSM; Lakeshore Model 7450). ND experiments were carried out at the Dhruva reactor of Bhaba Atomic Research Center, Trombay, employing the powder diffractometer-3 ($\lambda$ = 1.48 \AA).

\subsection{\label{sec:ExpStruct} Structural properties}
\begin{figure}[htp]
  \begin{center}
    \resizebox{85mm}{!} {\includegraphics *{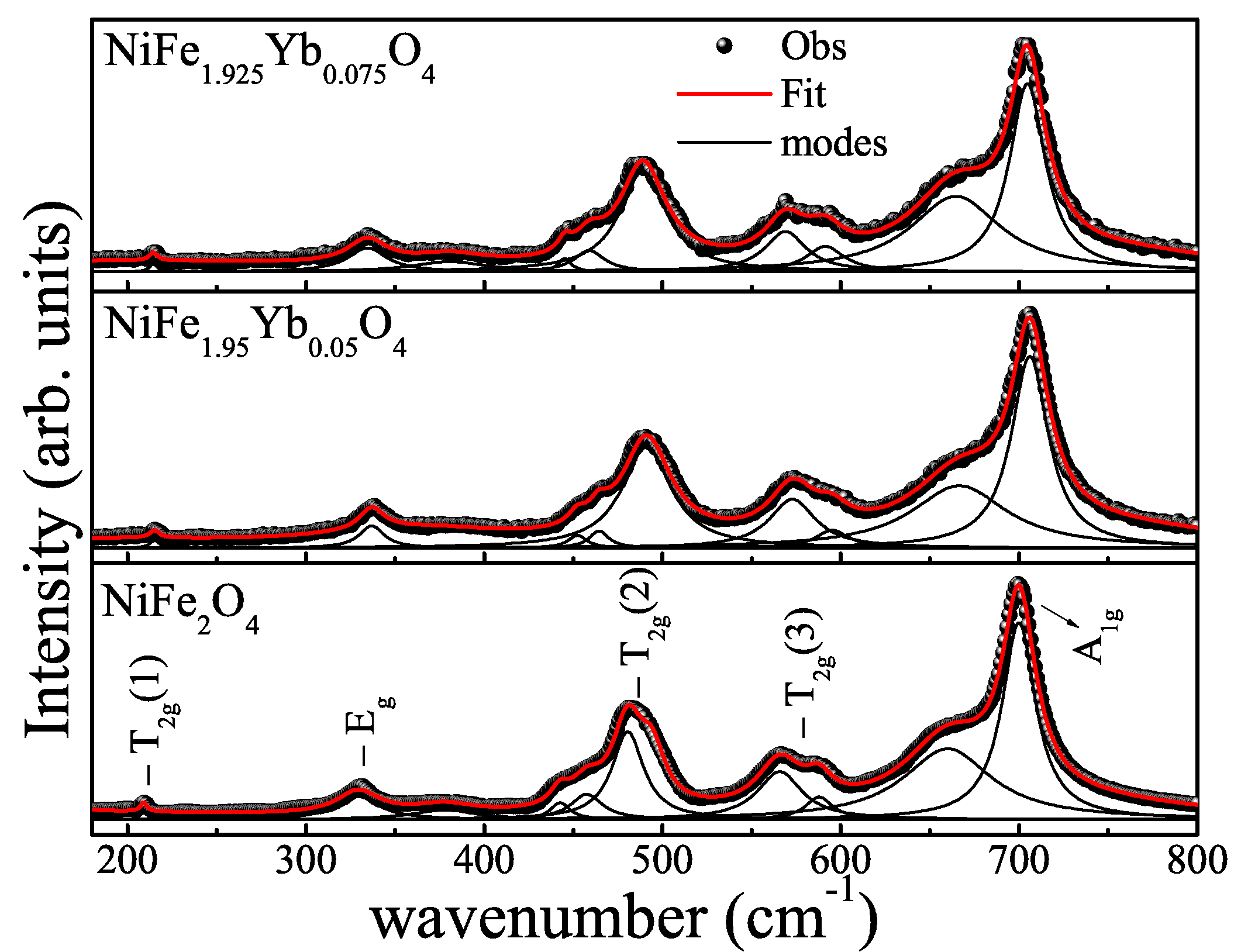}}
    \caption {Raman modes of NiFe$_{2-x}$Yb$_{x}$O$_{4}$ ($x$ = 0, 0.05, 0.075), confirm the single phase inverse spinel structure.}
  \label{fig:Raman}
  \end{center}
\end{figure}

The structural details, listed in Table~\ref{tab:Tab-Struct}, are obtained through Rietveld refinement using the General Structure Analysis System (GSAS) program. As expected, it is found that for all doping concentrations, the compounds were crystallized in the cubic inverse spinel phase (Fd$\bar{3}$m).  However, orthorhombic YbFeO$_3$ appears as the secondary phase. The weight fractions, ($i.e.$ phase fractions of different phases present in the compound) of inverse spinel phase and the secondary phase respectively are found to be 0.95 and 0.05 for \textit{x} = 0.05 and 0.91 and 0.09 for \textit{x} = 0.075. 
\vspace{6pt}
\\
The conventional unit cell of NFO, comprises of 8 divalent cations (Ni$^{2+}$), 16 trivalent cations (Fe$^{3+}$) and 32 oxygen anions (O$^{2-}$). There are two cationic sites in spinel structure: (i) tetrahedrally coordinated A-site (T$_d$ symmetry with Wyckoff position 8a) and (ii) octahedrally coordinated B-site (O$_h$ symmetry with Wyckoff position 16d). In inverse spinel NFO, A-sites are occupied Fe$^{3+}$ cations, and B-sites are  occupied by both Ni$^{2+}$ and Fe$^{3+}$ cations with equal distributions. \vspace{6pt}
\vspace{6pt}
\\
For Yb doped NFO, the Rietveld refinement was carried out by providing Yb occupancy in place of Fe in both A-site and B-site. The $\chi^2$ value for A-site occupancy is found to be greater than 3 for both the doping concentration whereas it is less than 2 for B-site occupancy. This suggests that Yb$^{3+}$ ion prefers to replace octahedral Fe$^{3+}$ ion. The replacement of Ni$^{2+}$ by Yb is not favored as the XPS studies, discussed next, suggest 3+ charge state to Yb. Table~\ref{tab:Tab-Struct} shows that, in agreement with earlier reports \cite{Kamala,KamalaJPC,RezlescuJPCM,RezlescuSSU}, an increment in the lattice constant was observed for the doped compounds which is attributed to the larger ionic radius of Yb$^{3+}$ (0.86 \AA) compared to that of  Fe$^{3+}$  (0.63 \AA).  Also from Table~\ref{tab:Tab-Struct}, it is clear that the octahedral (O-B) bond length has increased and tetrahedral (O-A) bond length has decreased with Yb substitution.  This implies that the Yb$^{3+}$ ions have replaced the Fe$^{3+}$ ions at octahedral site in the process of doping.
\begin{widetext}
\begin{center}
\begin{table}[htpb]
    \caption{Structural parameters of NiFe$_{2-x}$Yb$_{x}$O$_{4}$ ($x$ = 0, 0.05, 0.075). Here A and B represent the tetrahedral and octahedral sites respectively.}
    \label{tab:Tab-Struct}
\begin{tabular}{l l r @{.} l r @{.} l r @{.} l }
\hline
 Composition&  &\multicolumn{2}{c}{$x$ = 0} 	&\multicolumn{2}{c}{$x$ = 0.05}  &\multicolumn{2}{c}{$x$ = 0.075}	 \\
\hline
 \multicolumn{2}{l}{$\chi^{2}$ (goodness of fit) (B-site)}		&\multicolumn{2}{c}{1}	&1&13  &1&21\\
\multicolumn{2}{l}{$\chi^{2}$ (A-site)}		&\multicolumn{2}{c}{1}	&4&73  &3&20
\\
\multicolumn{2}{l}{$\omega_{rp}$ (weighted refined parameter)}		&1&68\% &1&60\% &1&99\% \\ \vspace{2 mm}
Lattice constant (\AA)	\cite{Ugendar}&	&8&341(5)   &8&343(6) &8&346(3) \\
\multirow{2}{*}{Weight fractions \cite{Ugendar}}   &Inverse Spinel &\multicolumn{2}{c}{1}  &0&95   &0&91   \\ \vspace{2 mm}
 &Secondary Phase (YbFeO$_3$)   &\multicolumn{2}{c}{0}  &0&05   &0&09       \\
\multirow{2}{*}{Bond length(\AA)}	&O$^{2-}$ - Fe$^{3+}$ (B)   & 2&032   &2&043    &2&075 \\ \vspace{2 mm}
 &O$^{2-}$ - Fe$^{3+}$ (A)	&1&9    &1&882  &1&827 \\ 
\multirow{4}{*}{Bond Angle (degrees)}   &Fe$^{3+}$(A) - O$^{2-}$ - Fe$^{3+}$/Yb$^{3+}$(B)  &123&11     &123&54 &124&82 \\
 &Ni$^{2+}$ - O$^{2-}$ - Ni$^{2+}$	&93&00	&92&42  &90&62 \\
 &O$^{2-}$ - Fe$^{3+}$ (A) - O$^{2-}$   &109&47 &109&47 &109&47\\
 &O$^{2-}$ - Fe$^{3+}$ (B) - O$^{2-}$   &93&08  &92&47  &90&62 \\
\hline
\end{tabular}
\end{table}
\end{center}
\end{widetext}
In order to confirm that the compounds formed in inverse spinel structure and that no other phases such as NiO or $\alpha$ -Fe$_2$O$_3$ or Fe$_3$O$_4$  are present, the materials were further investigated by taking Raman spectra.  Fig.~\ref{fig:Raman} shows the room temperature Raman spectra recorded in the wave number range 180 to 750 cm$^{-1}$. Five Raman modes \textit{viz.}, A$_{1g}$ (1), E$_{g}$ (1), T$_{2g}$ (3) corresponding to the inverse spinel phase with space group (Fd$\bar{3}$m) \cite{Ivanov, Iliev} were identified and indicated in Fig.~\ref{fig:Raman}.  Additional peaks, observed in all the compounds, are attributed to the presence of short-range ordering of Ni$^{2+}$ and Fe$^{3+}$ ions at the B-site \cite{Ivanov, Iliev, Fritsch}.
\subsection{\label{sec:ExpXPSND} Charge state and local spin moments: XPS and ND studies}
\begin{figure}[htpb]
  \begin{center}
    \resizebox{70mm}{!} {\includegraphics *{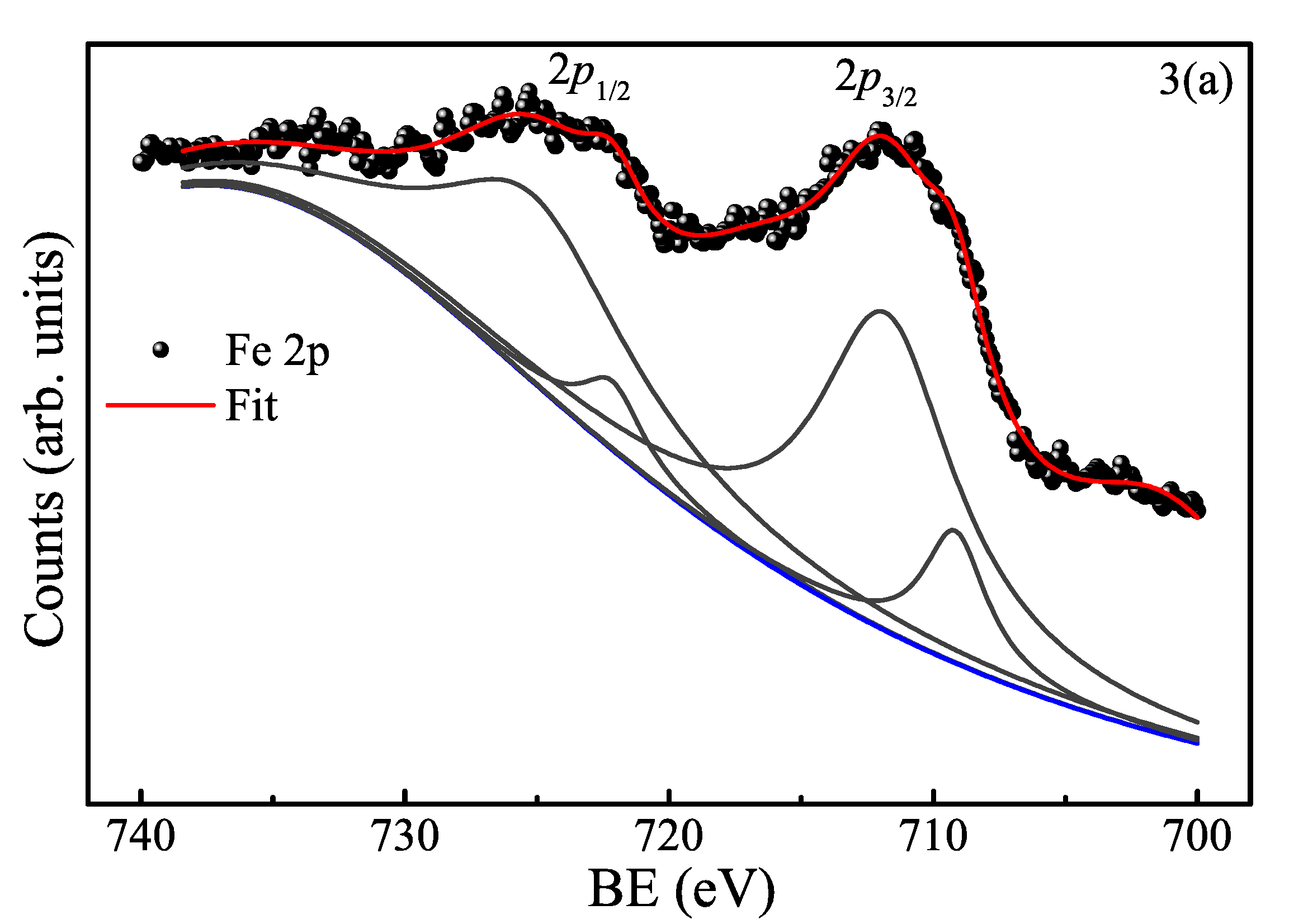}}
    \resizebox{70mm}{!} {\includegraphics *{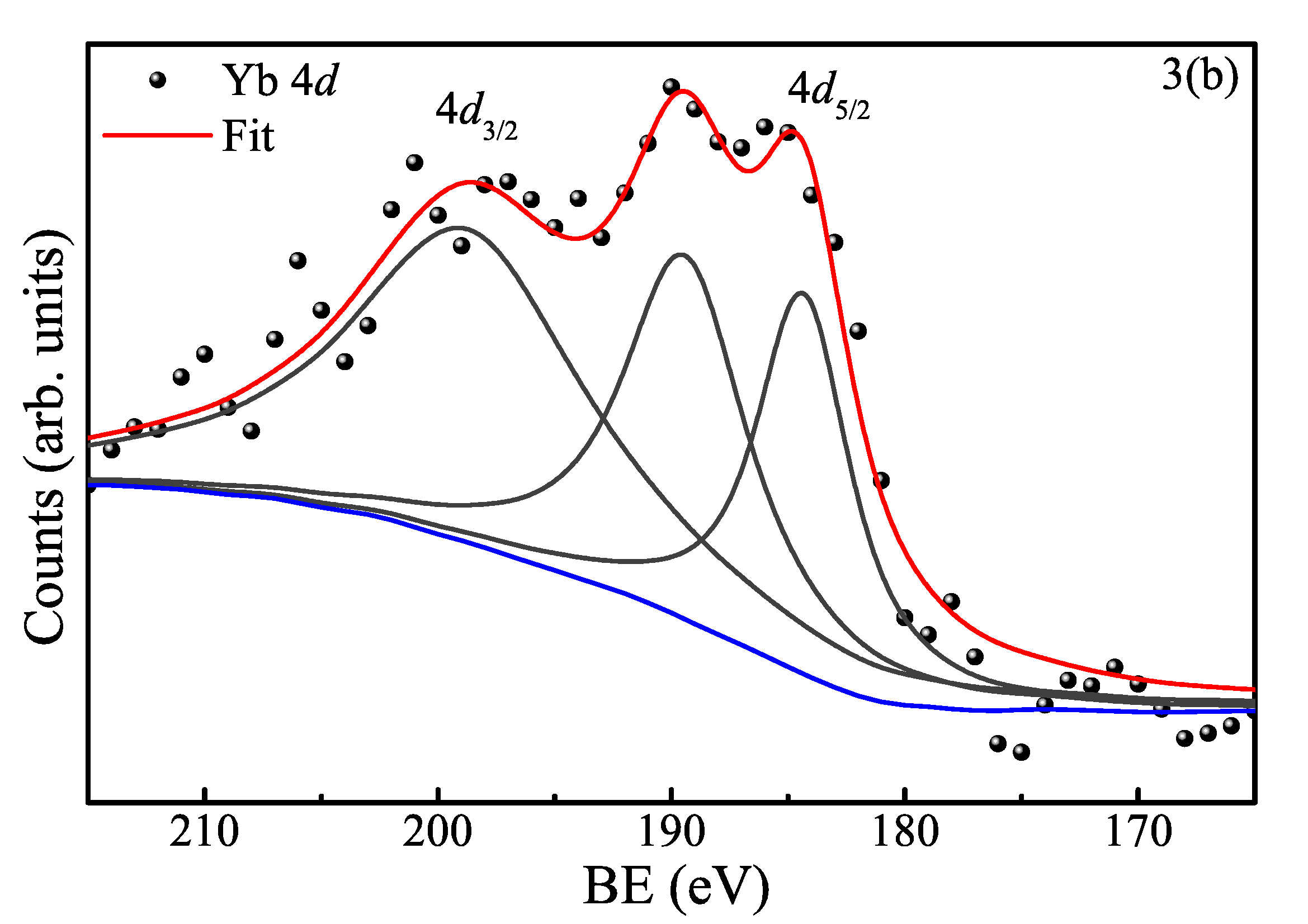}}
    \caption {Fe 2\textit{p} and Yb 4\textit{d}  core level XPS of NiFe$_{1.925}$Yb$_{0.075}$O$_{4}$. }
  \label{fig:XPS}
  \end{center}
\end{figure}
While majority of the lanthanide elements stabilize in +3 charge state, Yb stabilizes both in +2 and +3 charge states \cite{Jin-Seok, Yasui}.  In order to confirm the charge state of Yb and other cations, XPS measurements were carried out.  The Fe 2\textit{p} and Yb 4\textit{d} spectra of NiFe$_{1.925}$Yb$_{0.075}$O$_{4}$ are shown in Figs.~\ref{fig:XPS} (a) and (b), respectively.  The Fe 2\textit{p} spectrum consists of spin-orbit-split 2\textit{p}$_{3/2}$ and 2\textit{p}$_{1/2}$ peaks.  The binding energies (BEs) corresponding to both these peaks are in good agreement with literature \cite{Droubay, Stefan}.  The peaks also consist of doublets, which were attributed to Fe$^{3+}$ and Fe$^{2+}$ ions.
\vspace{6pt}
\\
\begin{figure}[ht]
  \begin{center}
    \resizebox{85mm}{!} {\includegraphics *{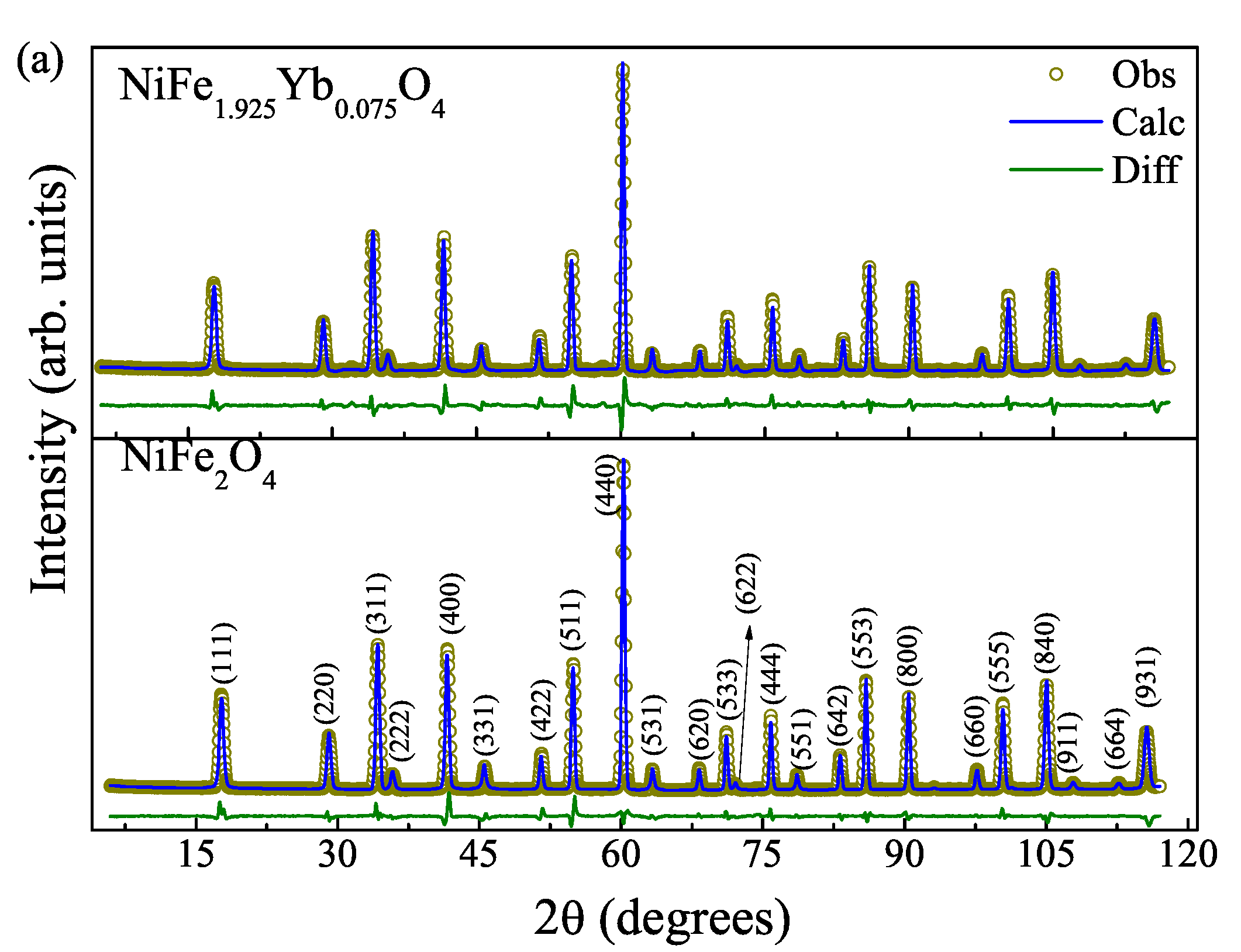}}
    \resizebox{85mm}{!} {\includegraphics *{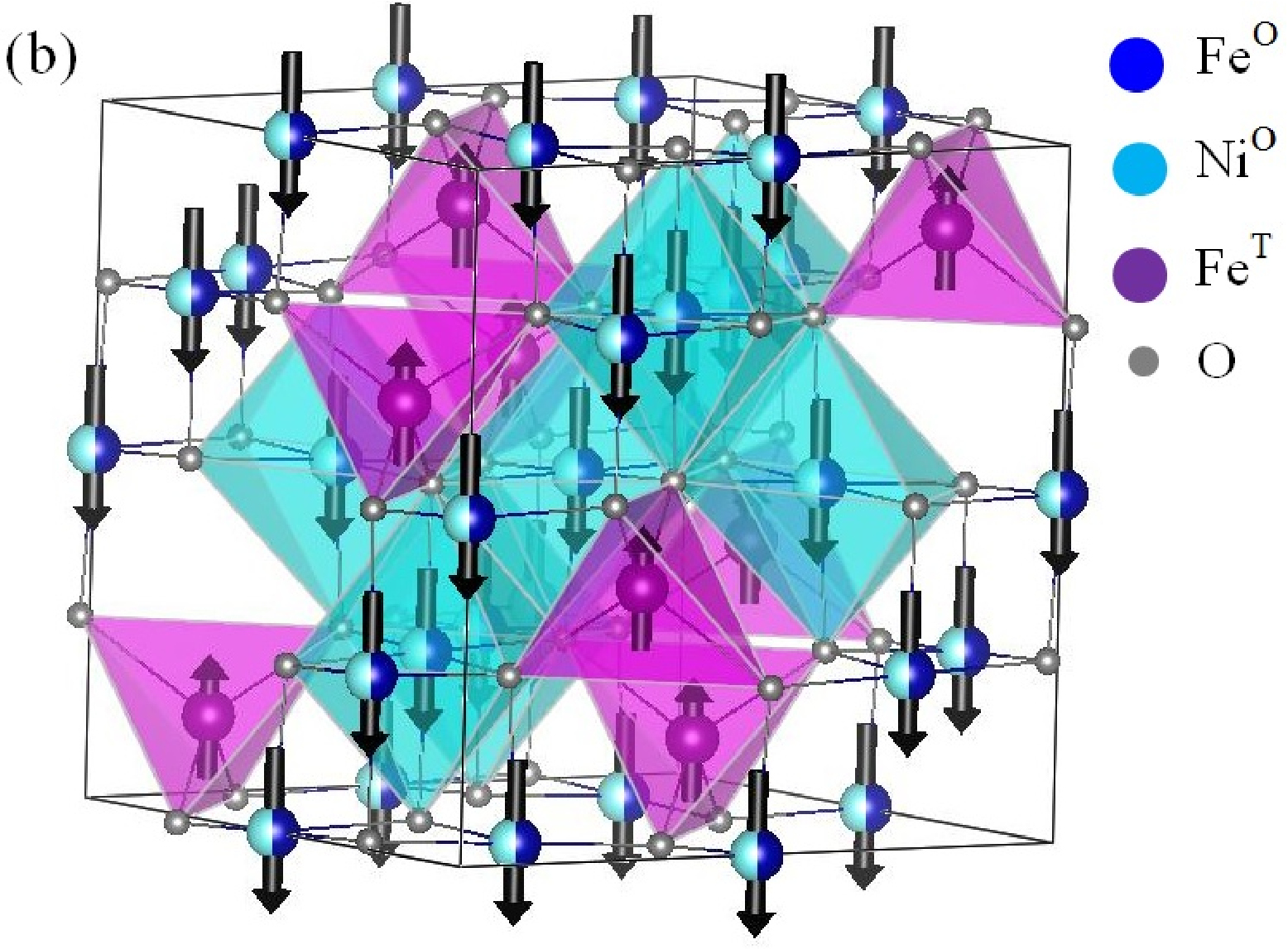}}
    \caption {(a) Indexed Rietveld refined ND pattern of NiFe$_{1.925}$Yb$_{0.075}$O$_{4}$ at 300 K. (b) Magnetic structure confirming the ferrimagnetic nature.}
  \label{fig:NDS}
  \end{center}
\end{figure}
The high BEs of 712 eV (in 2\textit{p}$_{3/2}$ peak) and 726.7 eV (in 2\textit{p}$_{1/2}$ peak) correspond to Fe$^{3+}$ ions.  The BEs 709.3 eV (in 2\textit{p}$_{3/2}$ peak) and 722.5 eV (in 2\textit{p}$_{1/2}$  peak) correspond to Fe$^{2+}$ ions.  This suggests that there exists a minor fraction of Fe$^{2+}$ ions in the synthesized sample. It may be noted that Fe BEs corresponding to +2 and +3 charge states are 709.2 eV and 711.2 eV respectively \cite{Stefan}.  Droubay \textit{et.al} \cite{Droubay}, by analyzing  Fe 2\textit{p} XPS spectra of $\alpha$ -Fe$_2$O$_3$ films and Fe 2\textit{p} (L-edge) X-ray absorption spectra of Fe$_2$O$_3$ (in the form of pressed pellet), have reported similar BEs which they attribute to the 2$p^5$3$d^6$L final state of Fe$^{3+}$ ions.
\vspace{6pt}
\\
The XPS spectrum of Ni 2\textit{p} (not shown)  exhibits spin-orbit-split 2\textit{p}$_{3/2}$ and 2\textit{p}$_{1/2}$ peaks at ~855 eV and ~874 eV respectively.  These binding energies are comparable to that of octahedral Ni$^{2+}$ found in NiO \cite{Stefan, Fujimori, Grosvenor}, suggesting the +2 charge state of Ni, in the compound investigated.  The XPS of Yb 4\textit{d} is shown in Fig.~\ref{fig:XPS} (b).  The two peaks centered at 184.5 eV and 199 eV are assigned respectively to 4\textit{d}$_{5/2}$ and 4\textit{d}$_{3/2}$ spin-orbit-split 4\textit{d} levels.  These values are in good agreement with that of Yb 4\textit{d} level binding energy in Yb$_2$O$_3$ \cite{Signorelli} which implies +3 charge state for Yb \cite{Signorelli}.  The spin-orbit-split was found to be 14.5 eV, which further confirms the +3 charge state for Yb.  Had it been in +2 state, the separation would have been 8.8 eV, as suggested by Hagstr\"{o}m \textit{et.al} \cite{Hagstrom}.  The O 1\textit{s} spectrum (not shown) consists of two peaks at 528.8 eV and 531.9 eV, which are ascribed to O$^{2-}$ ions at the octahedral coordinates and tetrahedral coordinates respectively.
\vspace{6pt}
\\
In order to understand the magnetic ordering and to quantify the local spin moments in NiFe$_{2-x}$Yb$_x$O$_4$, ND measurements were carried out. The ND patterns recorded at room temperature are shown in Fig. ~\ref{fig:NDS}(a). Rietveld refinement was done using the FULLPROF program. The site positions used were: A-site (Fe$^{3+}$) at 8\textit{a} ($\frac{1}{8}, \frac{1}{8}, \frac{1}{8}$), B-site (Ni$^{2+}$, Fe$^{3+}$, Yb$^{3+}$) at 16\textit{d} ($\frac{1}{2}, \frac{1}{2}, \frac{1}{2}$),  and oxygen at 32\textit{e} (0.26, 0.26, 0.26).  Thompson-Cox-Hastings pseudo-Voigt function was employed for the peak profile.  All the peaks were reconciled the cubic inverse spinel structure (\textit{Fd$\bar{3}$m}) and no impurity phase was observed. We would like to note that, as ND scattering factor is independent of atomic number (Z) the impurities phases with very low phase fraction may not be detected with this technique. However, if the impurity phase has heavier element as in the case of YbFeO$_3$, it can be detected in XRD measurement where the scattering factor increases with Z. Both the crystal and magnetic refinement parameters are listed in Table~\ref{tab:ND-ref-parameters}.
\vspace{6pt}
\\
From Table~\ref{tab:ND-ref-parameters}, it can be noted that the magnetization decreases upon substitution of Yb$^{3+}$. This decrease in magnetization is due to the smaller value of the magnetic moment of Yb$^{3+}$ (0.86 $\mu_B$) compared to that of Fe$^{3+}$ ion (5 $\mu_B$). This behaviour is in agreement with the literature which was obtained by M-H curves and M\"{o}ssbauer studies on NiFe$_{2-x}$Yb$_{x}$O$_{4}$ (\textit{x} = 0, 0.05 and 0.075) \cite{Ugendar}. In addition, the substitution of  rare earth ions (possessing high magnetic moment compared with Fe$^{3+}$) such as R$^{3+}$ (R = Dy, Gd, Sm, Ho) in NFO also caused a decrease in magnetization, which has been attributed to antiparallel alignment of R$^{3+}$ (R = Dy, Gd, Sm and Ho) to octahedral Fe$^{3+}$  moments \cite{Kamala, KamalaJPC, KamalaJAP}. 
\vspace{6pt}
\\
However, in the present case,
the moments within the B-site, \textit{viz}., Yb$^{3+}$, Fe$^{3+}$ and Ni$^{2+}$ moments, are parallel (Table~\ref{tab:ND-ref-parameters}).  In addition, the moments of Fe$^{3+}$ ions at the A-site and that of B-site moments (Ni$^{2+}$, Fe$^{3+}$ and Yb$^{3+}$) are antiferromagnetically aligned and the latter possessing higher moment than the former confirms the ferrimagnetic ordering. The magnetic structure determined from the ND measurements is presented in Fig.~\ref{fig:NDS} (b).
\vspace{6pt}
\\
The magnetization, measured at a field of 100 Oe, in the temperature range 300 - 900 K is shown in Fig.~\ref{fig:MT-Curie}.
The Curie temperature was extracted from inflection point (in M vs. T) which gives a minimum in $\frac{dM}{dT}$ vs. T at T = T$_C$ (shown in Fig.~\ref{fig:MT-Curie}(b)). The T$_C$ of NiFe$_{2-x}$Yb$_{x}$O$_{4}$ compounds were found to be 850 K, 837 K and 836 K respectively for \textit{x} = 0, 0.05 and 0.075.  For the undoped compound, our result is in close agreement with the reported value of T$_C$ = 853 K \cite{Chikazumi, Kamala}.   With Yb substitution in NFO, the T$_C$ drops from  850 K to 836 K.  In the subsequent section, the strengths of various  magnetic exchange interactions in this compound are examined, to understand the reason for the decrease in T$_C$. 
\begin{widetext}
\begin{center}
\begin{figure}[hptb]
\includegraphics[scale=0.45]{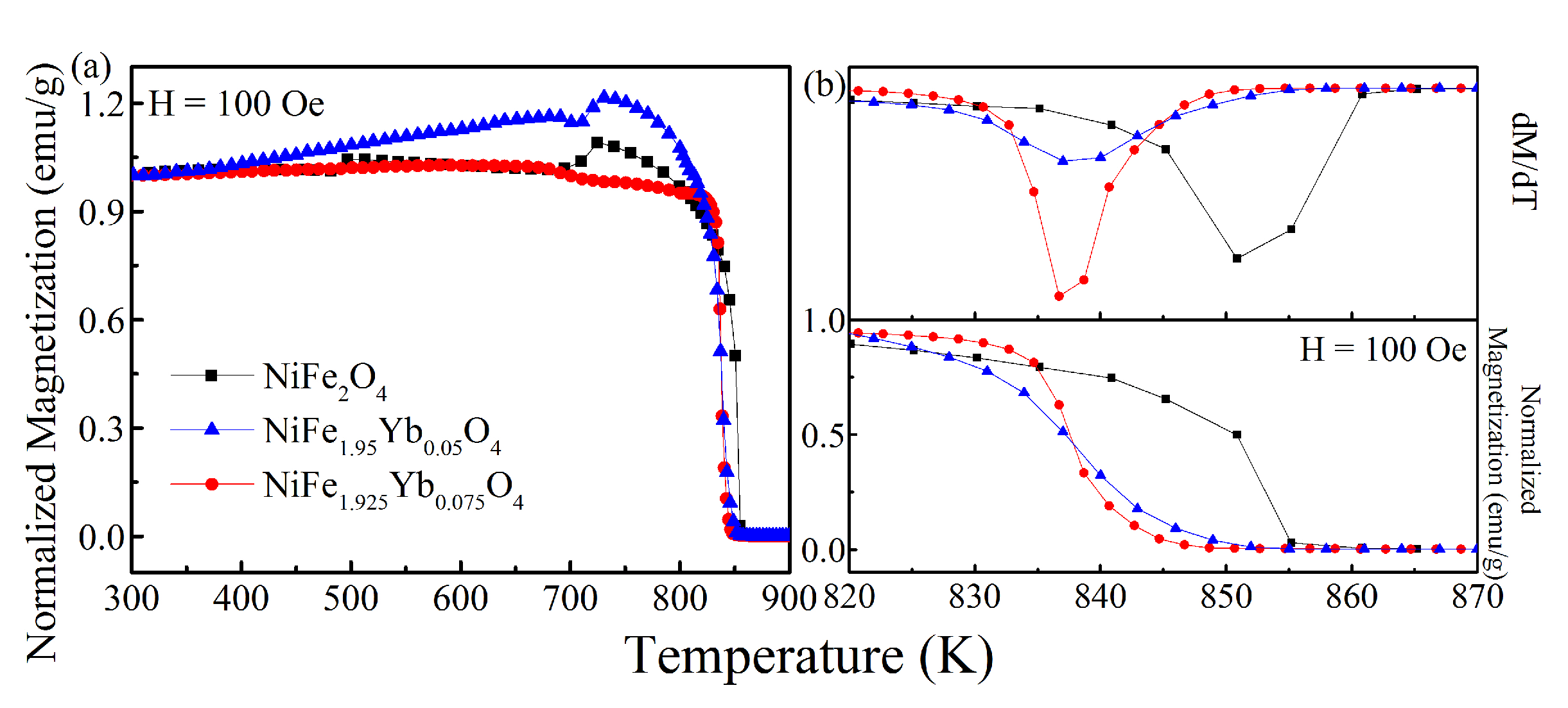}
\caption {\label{fig:MT-Curie} (a) Magnetization as function of temperature for NiFe$_{2-x}$Yb$_{x}$O$_{4}$ ($x$ = 0.05, 0.075). (b)  Top: $\frac{dM}{dT}$ vs. T near T$_C$. Bottom: M vs. T near  T$_C$. The inflection point (in M vs. T) implies a minimum in $\frac{dM}{dT}$ vs. T at T = T$_C$, which  decreases with substitution of Yb.}
\end{figure}
 \begin{table}[ht]
    \caption{Structural and magnetic parameters of NFO and  NiFe$_{1.925}$Yb$_{0.075}$O$_{4}$ as obtained from ND studies.}
    \label{tab:ND-ref-parameters}
    \begin{tabular}{l l l r @{.} l  r @{.} l}
\hline
 Compound  &	&  &\multicolumn{2}{l}{NFO}  &\multicolumn{2}{l}{NiFe$_{1.925}$Yb$_{0.075}$O$_{4}$}	 \\
 \hline
Lattice constant (\AA)	&    &  &8&3402(3)	&8&3446(1) \\
 &$\chi ^2$   &  &1&45	&1&71   \\
Bragg-R &Bragg-R &  &5&29	&3&85   \\ 
factors &R$_f$ factor &  &3&20	&3&58   \\ \vspace{2 mm}
 &Mag-R &  &9&95	&8&58   \\ 
Reliability	&R$_{p}$    &   &5&37	&6&79 \\ 
factors (\%)&R$_{wp}$    &   &7&87	&8&92 \\ \vspace{3mm}
&R$_{exp}$   &   &6&54	&6&84 	\\ 	\vspace{2mm}
&A-site &Fe$^{3+}$      &2&64	&2&56    \\  
 Magnetic     &\multirow{4}{*}{B-site} &Fe$^{3+}$      &-4&32	&-4&15    \\
 moments &   &Ni$^{2+}$      &-1&22	&-0&88    \\  
($\mu_B/f.u.$)&   &Yb$^{3+}$      &\multicolumn{2}{l}{-}	&-0&06 $^\S$   \\ \vspace{2 mm}
&   &Net      &-5&54	&-5&10    \\
\multicolumn{3}{l}{Net moment $\left| \mu_B - \mu_A \right| $}    &2&90	&2&54    \\
\hline
\multicolumn{7}{l}{\S As each f.u. contains 7.5\% of Yb, the moment per Yb ion is 0.86 $\mu_B$.}
\end{tabular}
\end{table}
\end{center}
\end{widetext}
\section{\label{sec:DFT-Stds}ELECTRONIC STRUCTURE FROM DFT STUDIES}
To study the formation of Yb impurity spin and its interaction with the host magnetic ordering and thereby to understand the origin behind the experimental observation of decrease in magnetization and T$_C$,  in this section DFT calculations were carried out and the results are presented in the subsequent sections.

\subsection{\label{sec:Comp-dtls}Computational details}

The \textit{ab initio} calculations are performed using the full potential linearized augmented plane wave method with local orbital basis (FP-LAPW + lo), as implemented in WIEN2k \cite{Blaha}. For NFO, there are three possible Ni/Fe cation distributions at the B-sites $viz.$ P$\bar{4}$m2, P4$_1$22, and Imma \cite{Ederer2012}. However, it has been found that while the P$\bar{4}$m2 configuration has higher energy, both P4$_1$22 and Imma configurations are lower in energy and almost have same ground state \cite{Fritsch}. Since the primitive unitcell of P4$_1$22 configuration consists of four formula unit compared to two formula unitcell for Imma, the latter was used for calculations \cite{Nuala}. As mentioned earlier, the octahedral sites are occupied equally by Fe$^{3+}$ and Ni$^{2+}$ ions. Even though the distribution of these ions is random in an experimentally synthesized sample, for computation, ordered distribution with Ni$^{2+}$ and Fe$^{3+}$ ions occupying alternate octahedral sites, was considered. The experimental structure was further optimized to study the ground state electronic and magnetic properties. 
\vspace{6pt}
\\
For the self-consistent calculations the plane wave cutoff RK$_{max}$ was taken to be 7.0 which yielded 3939 plane waves for the interstitial region. The muffin-tin radii of Ni, Fe, Yb, and O were taken as 1.88, 1.84, 1.97, and 1.58 a.u. respectively. For the computation of non-muffin tin matrix elements, L$_{max}$ was set to 4. The local orbitals included 4\textit{s} and 3\textit{d} states for Ni and Fe; 4\textit{f}, 6\textit{s} and 5\textit{d} for Yb and 2\textit{s} and 2\textit{p} states for O. Brillouin zone integration was performed using tetrahedron method on 8$\times$8$\times$8 k-grid. Out of the possible exchange correlation approximations, such as, LSDA \cite{Penicaud}, LSDA+SIC \cite{Szotek}, GGA \cite{Perron} and GGA+U \cite{Nuala} applied to NFO so far, the result with GGA+U matched well with the experimental magnetic ordering as well as band gap \cite{Sun:12}. Therefore, the results presented in this paper are obtained within the framework of GGA+U. The effective U (i.e. U - J = 3 eV) is applied to Ni-\textit{d}, Fe-\textit{d} and Yb-\textit{f} orbitals and the calculations are carried out using rotationally invariant Dudarev approach \cite{Dudarev}. Supporting the experimental observation, the calculations showed that Yb replacing the octahedral Fe (Fe$^{O}$) is favourable by 0.6 eV than Yb replacing the tetrahedral Fe (Fe$^{T}$).  Therefore, in rest of the article, the results for $\mathrm{Ni{Fe^{O}}_{1-x}Yb_xFe^{T}O_4}$ are presented.
\vspace{6pt}
\\
For the electronic structure calculations with substituents, a 2$\times$2$\times$2 supercell which gives rise to a sixteen \textit{f.u.} unit cell was constructed. One and two Yb atoms were substituted to construct NiFe$_{2-0.0625}$Yb$_{0.0625}$O$_{4}$ and NiFe$_{2-0.125}$Yb$_{0.125}$O$_{4}$ respectively. Even though these concentrations of Yb ions do not match exactly with those of the synthesized samples, NiFe$_{2-0.05}$Yb$_{0.05}$O$_{4}$ and NiFe$_{2-0.075}$Yb$_{0.075}$O$_{4}$, the qualitative features of Yb substitution are not expected to differ substantially. The optimized lattice constants of NiFe$_{2-x}$Yb$_x$O$_4$ are found to be 8.44\ \AA, 8.46\ \AA, and 8.53\ \AA \ for $x$ = 0, 0.0625 and 0.125, respectively. These agree well with experimental observation of lattice expansion with substitution of Yb. 

\subsection{\label{sec:ElcMag-dtls}Electronic and magnetic structure of pure and Yb doped NFO}
The electronic structure of pure NFO has been investigated by many in the past in the context of strong correlation effect \cite{Sun:12, Antonov}, spin-filter efficiency \cite{Nuala} and stability of normal and inverse spinel configuration \cite{Szotek, Perron, Fritsch}. In the present study, the electronic structure of NFO is revisited in order to examine the effect of substitution of Yb as well as to explain the insulating mechanism of the parent compound. 
\vspace{6pt}
\\
\begin{figure}[htpb]
    \begin{center}
    \resizebox{80mm}{!} {\includegraphics *{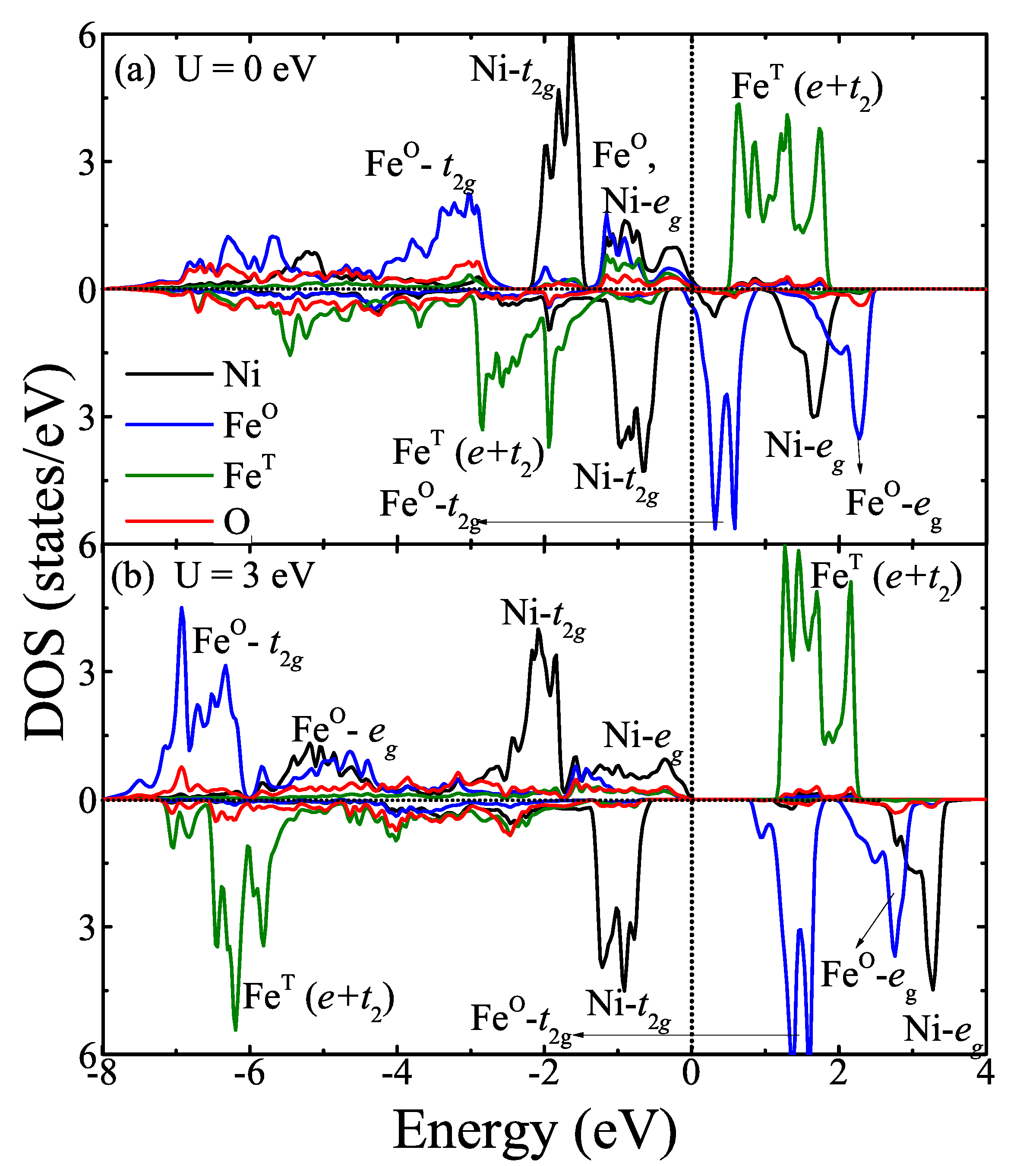}}
     \resizebox{80mm}{!} {\includegraphics *{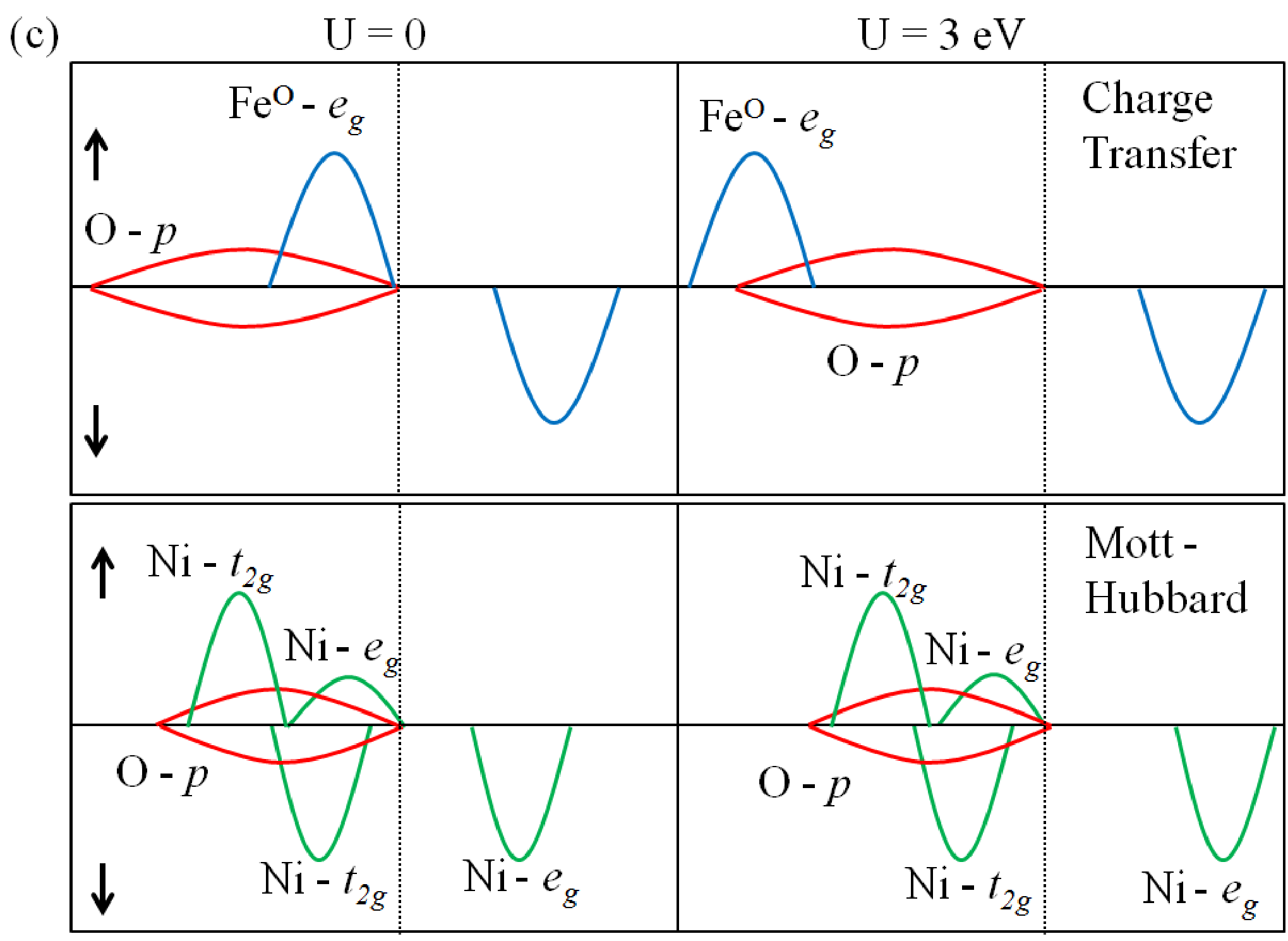}}
    \caption {Spin polarized DOS of NFO using (a) GGA and (b) GGA+U show that the correlation effect makes the system insulating. (c) Schematic diagram illustrating the insulating mechanism in this compound.}
  \label{fig:NF-DOS}
    \end{center}     
\end{figure}
In NFO, due to the octahedral crystal field effect of O-ligands, Ni and Fe$^{O}$-\textit{d} states are split into triply degenerate $t_{2g}$ and doubly degenerate $e_g$ states with the former lying lower in energy. Likewise, the tetrahedrally coordinated Fe-\textit{d} states are split into doubly degenerate \textit{e} and triply degenerate $t_2$, where the latter has higher energy. The densities of states in Fig.~\ref{fig:NF-DOS} show these crystal field effects. From the figure, it can also be seen that in the spin up channel, both $t_{2g}$ and $e_g$ states of Ni and Fe$^{O}$ are occupied. However, in the spin down channel, the Fe$^O$-\textit{d} states are empty and only Ni-$t_{2g}$ states are occupied. On the other hand, the spin-resolved occupancy of the  Fe$^{T}-d$ states is opposite to that of Fe$^{O}$.  These occupancies together confirm Ni$^{2+}$ and Fe$^{3+}$ charge states and also the antiparallel alignment of Fe$^{T}$ spins with Ni and Fe$^{O}$ spins. Such a spin alignment makes NFO ferrimagnetic with net magnetic moment of 2 $\mu_B$/f.u. (see Table~\ref{tab:DFTmoments}), which is in agreement with those reported in literature \cite{Kamala, KamalaJPC, Chikazumi, Ugendar, Ederer}.
\begin{table}[htp]
\begin{center}
\caption{
The local magnetic moment (estimated within the muffin-tin sphere) of Ni, Fe$^{O}$, Yb and Fe$^{T}$ of NiFe$_{2-x}$Yb$_x$O$_4$ (\textit{x} = 0, 0.0625 and 0.125) using GGA + U( = 3eV)}
\label{tab:DFTmoments} 
\begin{tabular}{l r r r }
\hline
	&\multicolumn{3}{c}{Moments ($\mu_B$/f.u.) }	\\
Composition   &\textit{x} = 0	&\textit{x} = 0.0625	&\textit{x} = 0.125  \\
\hline 
Ni		&1.57	&1.57	&1.58		\\
Fe$^{O}$	&3.98		&3.98		&4.07		\\
Fe$^{T}$	&-3.83 	&-3.83  	&-3.93	\\
Yb			&	-		&0.90		&0.90		\\
O			&0.06		&0.05		&0.04		\\
Total		&2.00		&1.75		&1.50		\\
\hline
\end{tabular}
\end{center}
\end{table}
\vspace{6pt}
\\
In order to examine the role of strong correlation effect on the insulating behavior, the partial densities of states obtained using GGA and GGA+U are shown in Fig.~\ref{fig:NF-DOS} (a) and (b) respectively. Since Ni-$t_{2g}$ states are occupied in both the spin channels, they are not affected by the on-site repulsion U. The half-filled Fe$^{O}-d$ states exhibit maximum strong correlation effect. Specifically, the Fe$^O-e_g$ spin-up states, which are at the Fermi level (E$_F$) along with O-\textit{p} states in GGA, are now pushed down in energy, with the inclusion of U. Since O-\textit{p} states occupy the E$_F$, it confirms the charge-transfer mechanism. However, the same is not the case for Ni-$e_{g}$ states. In the spin majority channel, the band center of Ni-$e_g$ lies above the band center of O-p in order to favor the Mott-Hubbard mechanism. The Fe$^{T}-d$ states are away from the E$_F$ both in GGA and GGA+U and hence, have negligible role in determining the insulating mechanism in this system. As a whole, the entire system behaves as mixed insulator falling under the Zaanen-Sawatzky-Allen classification scheme \cite{Zaanen}. The insulating mechanism in this compound is summarized schematically in Fig.~\ref{fig:NF-DOS} (c). We find that the mechanism remains valid for higher values of U.
\begin{figure}[htpb]
\begin{center}
     \resizebox{80mm}{!} {\includegraphics *{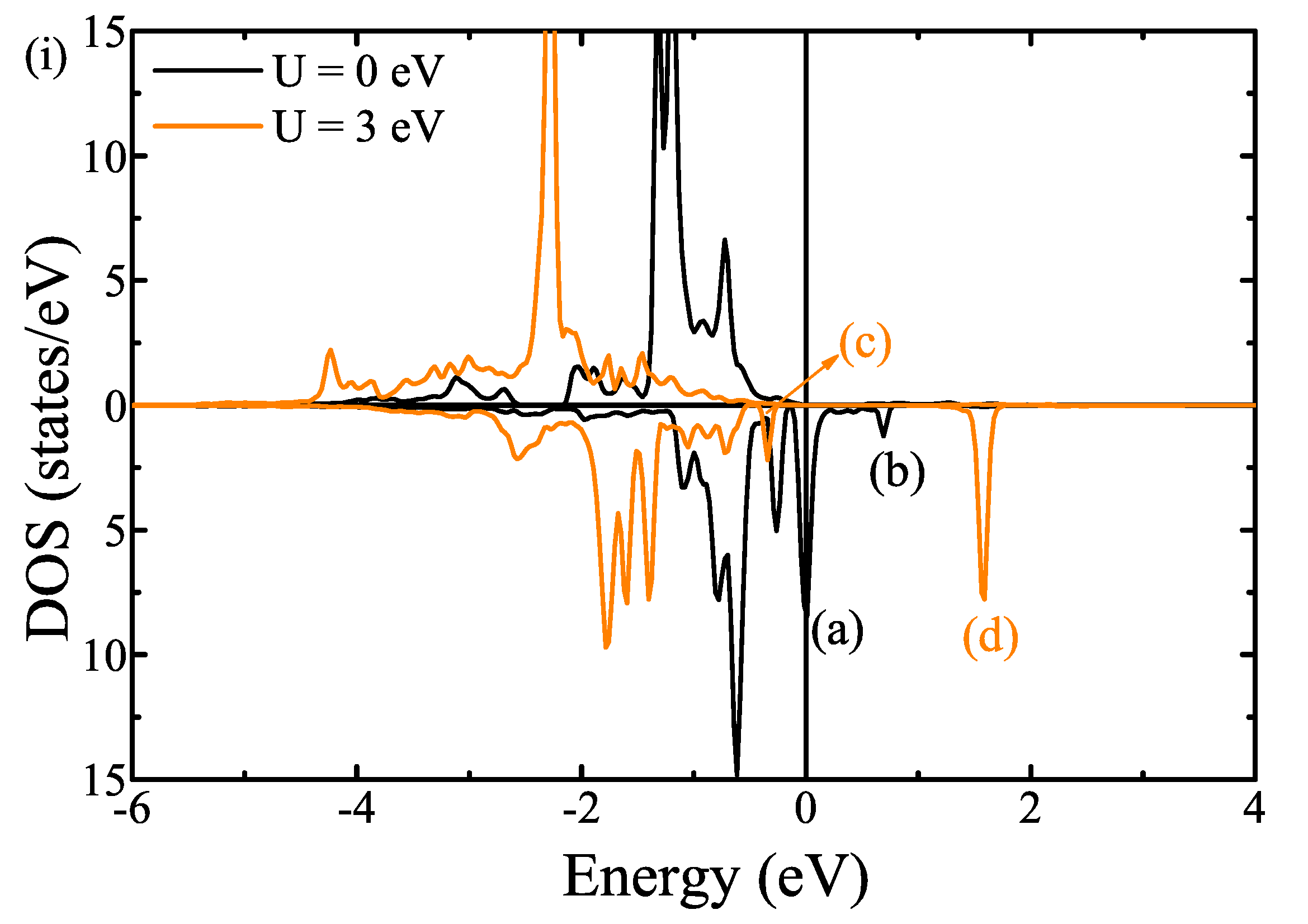}}
     \resizebox{80mm}{!} {\includegraphics *{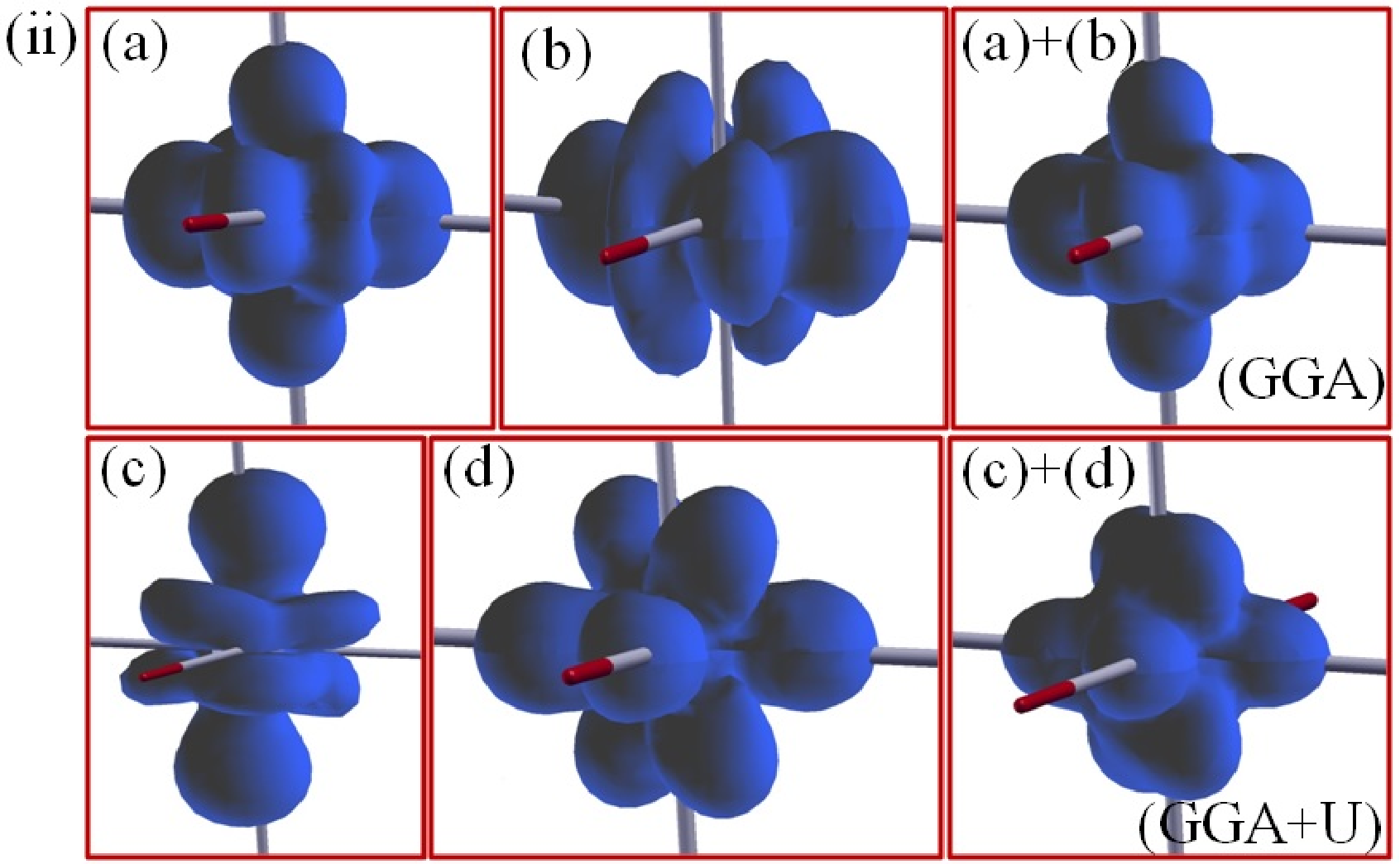}}
    \caption {(i) Partial Yb-\textit{f} DOS for NiFe$_{2-0.0625}$Yb$_{0.0625}$O$_4$. The partially occupied states (U = 0) split to form occupied lower Hubbard and unoccupied upper Hubbard bands under the influence of U. (ii) The electronic charge densities of the partially occupied states (U = 0) and the lone unoccupied states (U = 3eV) are also shown.}
  \label{fig:Yb-f-DOS}
 \end{center} 
 \end{figure}
 \\
With Yb substitution at the octahedral site, it was found that, both within GGA and GGA+U, Yb did not affect the Ni and Fe \textit{d}-states and hence, only the Yb-\textit{f} partial DOS is plotted in Fig.~\ref{fig:Yb-f-DOS}. The GGA only calculation shows two prominent peaks near E$_F$ which are denoted as (a) and (b). These peaks together form partially occupied \textit{f}-states and from the charge density plot, shown in Fig.~\ref{fig:Yb-f-DOS}, they are found to be linear combinations of $f_{z^3}$ and $f_{z(x^2 - y^2)}$ states. The rest of the $f$ orbitals lie lower in energy and are completely occupied. This can be understood from crystal field effect. As Yb$^{3+}$ is in an octahedral site and these $f$-orbitals are nearly along the axis of the YbO$_{6}$ octahedra, they experience stronger Coulomb repulsion by the O-ligands, compared to the rest and hence, lie higher in energy. 
\vspace{6pt}
\\
The strong correlation effect splits these partially occupied states to lower Hubbard band (LHB) and upper Hubbard band (UHB) which are also reflected in the DOS obtained with U = 3 eV. In Fig.~\ref{fig:Yb-f-DOS}(i) the peak (c) constitutes the LHB and the peak (d) constitutes the UHB. From the electronic density plot shown in the Fig.~\ref{fig:Yb-f-DOS}(ii), LHB is found to be of $f_{z^3}$ character and UHB is found to be of $f_{z(x^2 - y^2)}$ character. Since in the spin-majority channel Yb-$f$ states are completely occupied and in the spin-minority channel only $f_{z(x^2 - y^2)}$ is empty, this further reveals that Yb is in +3 charge state, confirming the experimental observation.
\subsection{ \label{sec:MagExch} Magnetic exchange interactions of pure and Yb substituted compound}
\begin{figure}[htpb]
     \includegraphics[scale=0.45]{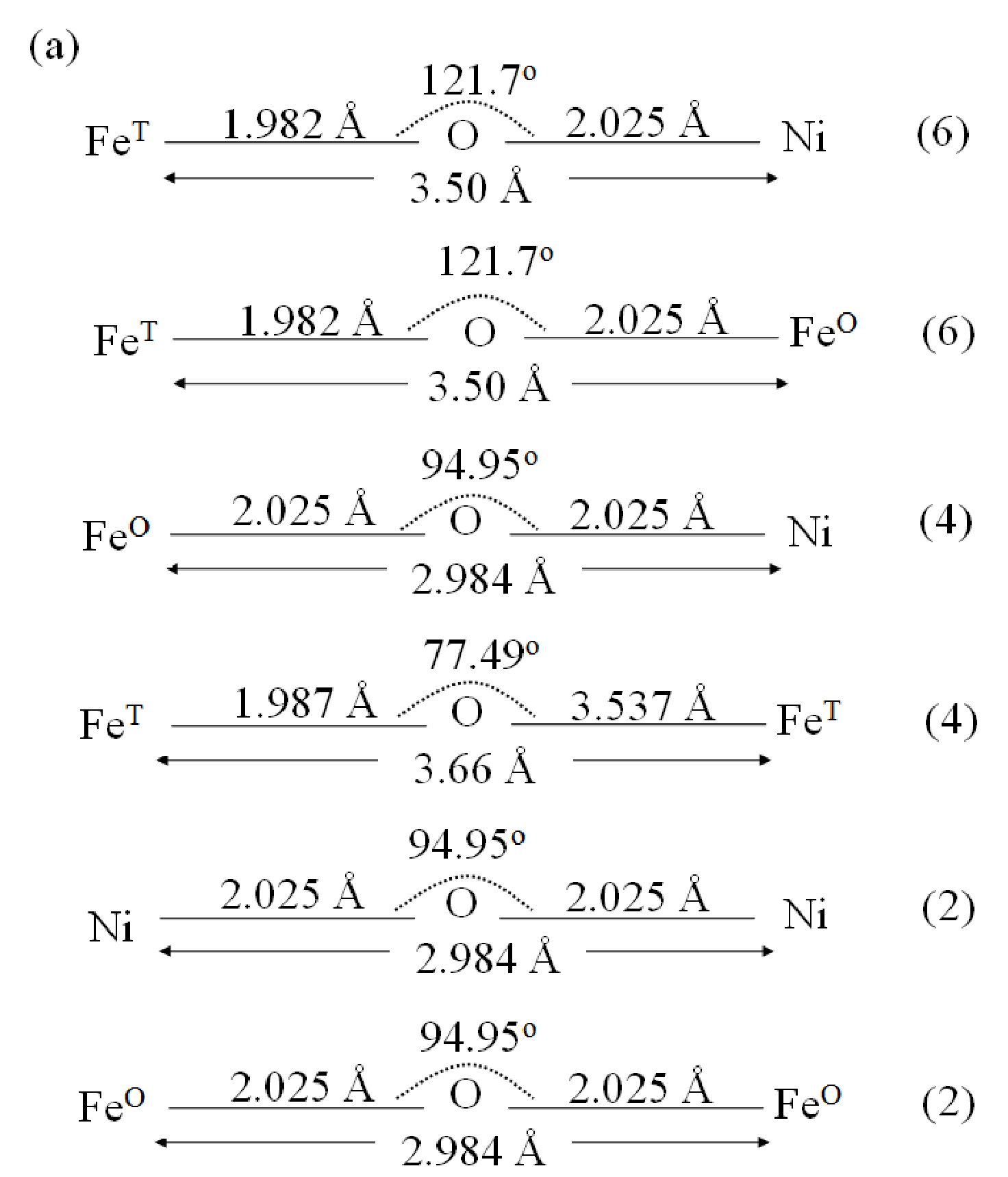}
     \resizebox{5mm}{!} { }
     \includegraphics[scale=0.45]{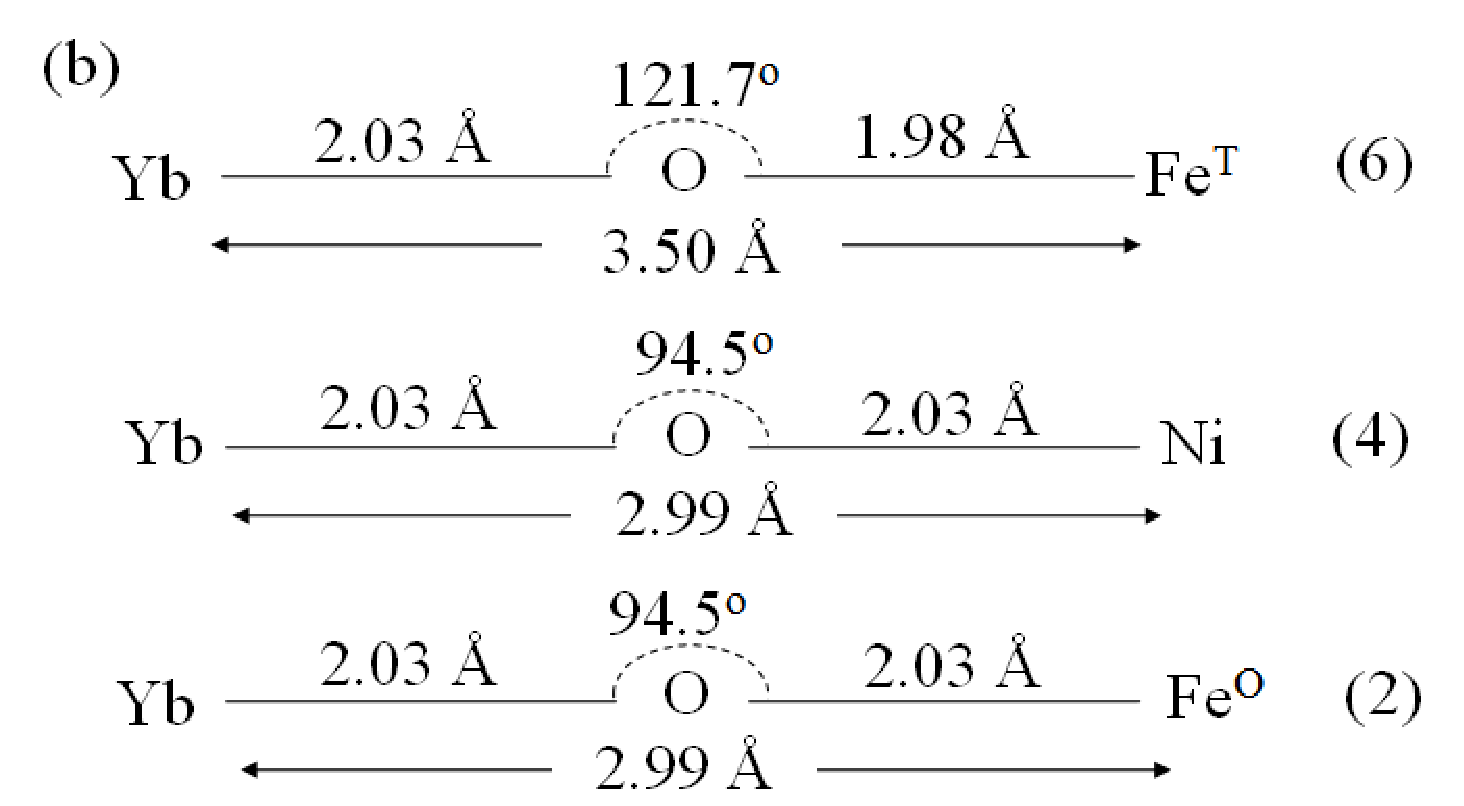}
    \caption {(a) Several exchange paths between the transition metal cations via the oxygen anion considered for NFO. The maximum cation-cation separation is restricted to 3.66 \AA. The larger exchange paths are ignored as the coupling strength of them is expected to be negligible. (b) For NiFe$_{2-0.0625}$Yb$_{0.0625}$O$_4$, only three different exchange paths are taken into account. The number written within the parenthesis (extreme right) implies number of such identical exchange paths from reference spin.}
  \label{fig:SD-NFO}
  \end{figure}
While studying the magnetization as a function of temperature in Fig.~\ref{fig:MT-Curie}, it was found that the Yb substitutions led to decrease in the Curie Temperature of the ferrimagnetic ordering. Since T$_C$ is determined from the spin-exchange interaction strengths J, it is imperative to identify the dominant interaction paths in the NFO and how they are affected by substitution of Yb. In this section, the Heisenberg Hamiltonian, $H$ = - $\sum_{i,j}$ J$_{ij}$ $\vec{S}_i \cdot \vec{S}_j$ was employed to NiFe$_{2-x}$Yb$_x$O$_{4}$ (\textit{x} = 0, 0.0625) compounds over several exchange paths between the transition metal cations, as shown in Fig.~\ref{fig:SD-NFO}. The value of J$_{ij}$ between two neighboring spins $S_i$ and $S_j$ of the dimer \textit{i-j} was obtained from the relation: J$_{ij} = E_{i\uparrow j\downarrow} - E_{i\uparrow j\uparrow}$. Here, E is the energy of the corresponding spin configurations and was evaluated from DFT calculations. Since multiple inter-coupled spin-dimers are involved in this system, the energies of several configurations were calculated to get the values of Js.
\vspace{6pt}
\\
For NFO, six exchange interaction paths, as shown in Fig.~\ref{fig:SD-NFO}(a), were considered and therefore, a minimum of seven magnetic configurations were necessary. Table~\ref{Tab:ExcInt} lists these seven configurations as well as the relation between the J$_{ij}$ and E of these configurations. Simultaneous solutions of these relations give the values of J$_{ij}$ which are given in Table~\ref{tab:Js-YbNFO}. The results are comparable to those obtained by Srivastava $et$ $al$ \cite{Srivastava} and by Cheng \cite{Chung}. The former has estimated J using Anderson transfer integral \cite{Anderson} as well three sublattice model. They have found that the strongest exchange interaction is J$_{Fe^OFe^T}$ which is in  agreement with our study. In addition, except J$_{NiNi}$, the signs of the exchange interactions are also found to be same.   Only three J values (J$_{NiFe^O}$, J$_{NiFe^T}$, J$_{Fe^TFe^O}$) have been estimated by Cheng \cite{Chung}, by pseudopotential based DFT calculations and the present results are in good agreement with them \cite{Chung}.
\vspace{6pt}
\\
Table~\ref{tab:Js-YbNFO} suggests that each of the exchange interactions is antiferromagnetic in nature (as shown in Fig.~\ref{fig:SD-YbNFO}). However, the ground state magnetic ordering shows that the spins at the octahedral sites (henceforth, referred as octahedral spins) are aligned parallel and there is an antiparallel alignment between the octahedral and tetrahedral spins. This can be explained from the schematic triangles (see Fig.~\ref{fig:SD-YbNFO}) representing the magnetic coupling between two octahedral and one tetrahedral spins. Had the octahedral spins been antiparallel (represented by the dashed arrow), one of them would have been parallel to the tetrahedral spin. Since the strength of antiferromagnetic exchange interaction between the neighboring octahedral and tetrahedral spins are far more stronger compared to that between neighboring spins (see the Fig.~\ref{fig:SD-YbNFO}), the octahedral spins are forced to align parallel to achieve the ground state magnetic ordering as shown by the solid arrows.
\vspace{6pt}
\\
To explain the effect of Yb spin-half impurity on the host magnetic ordering, the exchange interaction strengths J$_{YbFe^T}$, J$_{YbFe^O}$ and J$_{YbNi}$ along the paths shown in Fig.~\ref{fig:SD-NFO}(b) were estimated. The equations listed in the bottom part of Table~\ref{Tab:ExcInt} were used to estimate these magnetic interactions. Here, the effect of Yb$^{3+}$ on the host-host spin interactions was neglected. The estimated Yb-host J$_{ij}$s, are listed in Table~\ref{tab:Js-YbNFO}.
\vspace{6pt}
\\
From Table~\ref{tab:Js-YbNFO}, it is seen that $Yb-Fe^T$ coupling is significantly weak (-3.73 meV) in comparison to that of host octahedral-tetrahedral spin interactions: $Fe^O-Fe^T$ (-36.52 meV) and $Ni-Fe^T$ (-24.8 meV). At the same time, the $Yb-Fe^O$ (-23.72 meV) and $Yb-Ni$ (-23.05 meV) antiferromagnetic coupling have become stronger in comparison to that of host octahedral-octahedral spin interactions: $Fe^O-Fe^O$ (-19.12 meV) and $Fe^O-Ni$ (-4.07 meV).  Based on the coordination numbers of this inverse spinel structure, it was noted that there are four neighboring $Yb-Ni$ ($Fe^O-Ni$), two neighboring $Yb-Fe^O$ ($Fe^O-Fe^O$) and six neighboring $Yb-Fe^T$ ($Fe^O-Fe^T$) magnetic interactions. From the subtraction: (2J$_{YbFe^O}$+4J$_{YbNi}$+6J$_{YbFe^T}$)-(2J$_{Fe^OFe^O}$+4J$_{Fe^ONi}$+6J$_{Fe^OFe^T}$), it is found that there is a net decrease in the strength of the  antiferromagnetic interaction by 7 meV/f.u. in the NiFe$_{2-0.0625}$Yb$_{0.0625}$O$_4$ compound, which in turn decreases the T$_C$.
\begin{widetext}
\begin{center}
\begin{figure}[H]
\begin{center}
     \resizebox{100mm}{!} {\includegraphics *{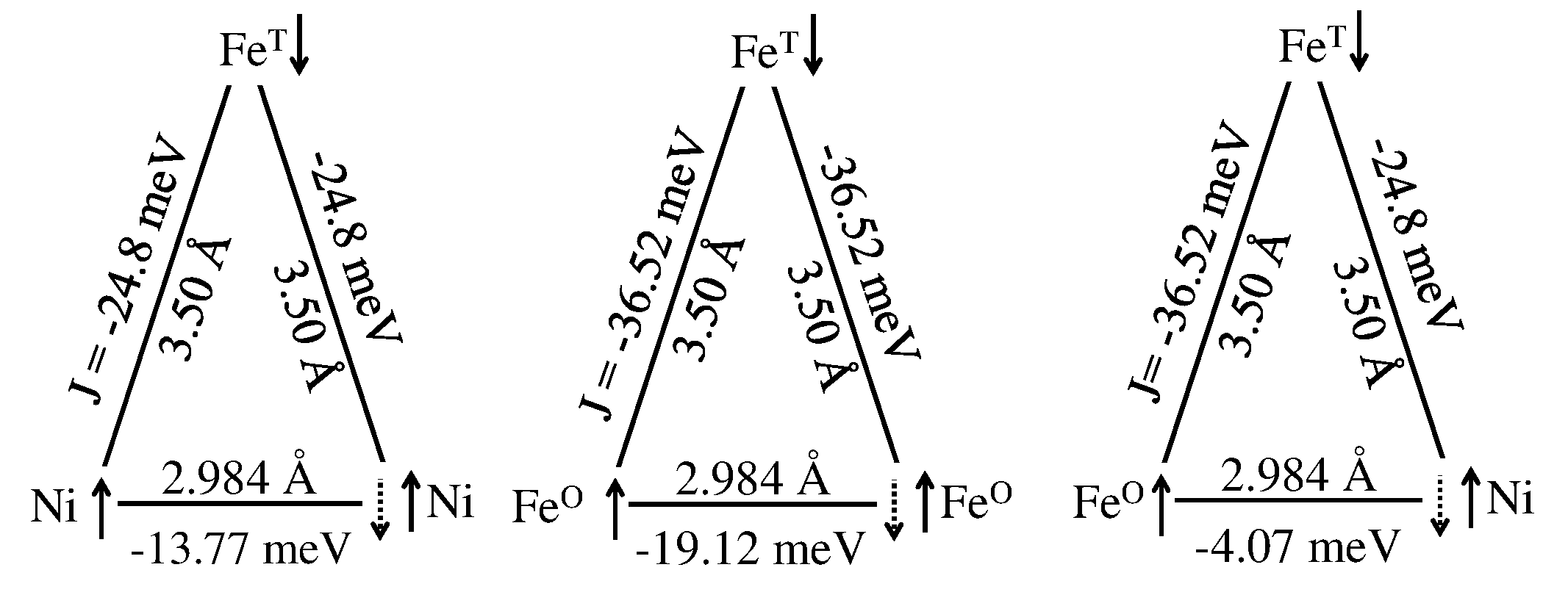}}
    \caption { Schematic of two neighboring octahedral spins join the tetrahedral spin to form a triangle. The solid arrows indicate the ground state magnetic ordering of that particular spin whereas the dotted one is suggested by the Js as calculated from the above equations.}
  \label{fig:SD-YbNFO}
  \end{center}
\end{figure}  
\begin{table}[htp]
\caption{(Pure) The seven magnetic configurations and the relation between the corresponding total energies and six exchange interactions (see Fig.~\ref{fig:SD-NFO}(a)) evaluated in this work. The equations are obtained by keeping the spin at one of the Fe$^O$ site (termed as reference spin) fixed and flipping the neighboring Ni, Fe$^O$ and Fe$^T$ spins as indicated in the table.  (Yb substituted) In the case of  Yb substitution, four necessary configurations are considered so that J$_{YbNi}$, J$_{YbFe^O}$ and J$_{YbFe^T}$ (see Fig.~\ref{fig:SD-NFO}(b)) can be evaluated. }
\label{Tab:ExcInt} 
\begin{tabular}{c c c c c c l}
\hline
& Configurations & Ref. spin & \hspace*{0.3cm}Flipped & spins &  & \hspace*{2.5cm} {Energy Difference} \\
\multicolumn{2}{c}{Pure (NFO)} & Fe$^{O}$ &Ni &Fe$^O$ &Fe$^T$\\
\hline
 & 1	 & $\uparrow$ &$\uparrow$    &$\uparrow$	&$\downarrow$ 	&$E_1$	(Ground state)	\\
 & 2	 & $\uparrow$ &$\downarrow$    &$\uparrow$	 &$\downarrow$	& $E_2 - E_1$ = 4J$_{NiNi}$ +  8J$_{NiFe^O}$ - 12J$_{NiFe^T}$		\\
& 3 & $\uparrow$ 	&$\uparrow$    &$\downarrow$	&$\downarrow$		&$E_3 - E_1 $ = 8J$_{NiFe^O}$ + 4J$_{Fe^OFe^O}$ - 12 J$_{Fe^OFe^T}$		\\
& 4 & $\uparrow$ 	&$\uparrow$    &$\uparrow$	&$\uparrow$	& $E_4 - E_1$ = - 12J$_{NiFe^T}$ - 12J$_{Fe^OFe^T}$	+ 8J$_{Fe^TFe^T}$	\\
& 5 & $\uparrow$ 	&$\downarrow$    &$\downarrow$	&$\downarrow$	&$E_5 - E_1$ = 4J$_{NiNi}$ + 12J$_{NiFe^O}$ - 12J$_{NiFe^T}$ + 4J$_{Fe^OFe^O}$ - 12J$_{Fe^OFe^T}$		\\
& 6	& $\uparrow$  &$\downarrow$    &$\uparrow$	 &$\uparrow$	&$E_6 - E_1$ = 4J$_{NiNi}$ + 8J$_{NiFe^O}$ - 20J$_{NiFe^T}$ -   12J$_{Fe^OFe^T}$  + 8J$_{Fe^TFe^T}$		\\
& 7 & $\uparrow$  &$\uparrow$    &$\downarrow$    &$\uparrow$    &$E_7 - E_1$ = 8J$_{NiFe^O}$ - 12J$_{NiFe^T}$ + 4J$_{Fe^OFe^O}$  - 20J$_{Fe^OFe^T}$ + 8J$_{Fe^TFe^T}$	 \\
\\
\multicolumn{2}{c}{Yb Substituted} & Yb &Ni &Fe$^O$ &Fe$^T$\\
\hline
 & 1 & $\uparrow$ & $\uparrow$    &$\uparrow$	&$\downarrow$ 	&$E_1^\prime$ (Ground state) \\
 & 2 & $\uparrow$ & $\downarrow$ & $\uparrow$ & $\downarrow$ & $E_{2}^\prime -E_{1}^\prime$ =  4J$_{NiNi}$ + 6J$_{NiFe^O}$ -12J$_{NiFe^T}$ + 2J$_{YbNi}$\\
& 3 & $\uparrow$ & $\uparrow$ & $\downarrow$ & $\downarrow$ & $E_{3}^\prime -E_{1}^\prime$ = 2J$_{Fe^OFe^O}$ + 8J$_{NiFe^O}$ - 12J$_{Fe^OFe^T}$ + 2J$_{YbFe^O}$\\
& 4 & $\uparrow$ & $\uparrow$ & $\uparrow$ & $\uparrow$ & $E_{4}^\prime -E_{1}^\prime$ =  -10J$_{Fe^OFe^T}$ -12J$_{NiFe^T}$ +8J$_{Fe^TFe^T}$ - 2J$_{YbFe^T}$\\
\hline
\end{tabular}
\end{table}
\begin{table}[ht]
\caption{The strength of exchange interactions J$_{ij}$ for pure and Yb substituted NFO. The values are obtained from the simultaneous solution of the equations of Table~\ref{Tab:ExcInt}. For comparison purpose the values from the literature are also listed here. Ref. \cite{Srivastava} has calculated J$_{ij}$ per unpaired spin at each site. Hence the comparison is made with S$_i$S$_j$ J$_{ij}$, where S$_i$ is the total number of unpaired spins at the {\it i}-th site.}
\label{tab:Js-YbNFO}
\begin{tabular}{m{1.5cm} m{2cm} r @{.} l r @{.} l r @{.} l r @{.} l m{2cm} r @{.} l}
\hline
Type of   &Interaction   &\multicolumn{8}{c}{J values in meV}
\\
Interaction	&Distance (\AA)	&\multicolumn{2}{l}{Present} &\multicolumn{2}{l}{Three sublattice} &\multicolumn{2}{l}{Anderson transfer}	&\multicolumn{2}{l}{Pseudopotential}	&Yb &\multicolumn{2}{l}{Present}	 
\\
&	&\multicolumn{2}{l}{work} &\multicolumn{2}{l}{model Ref \cite{Srivastava}}	&\multicolumn{2}{l}{integral \cite{Srivastava}} &\multicolumn{2}{l}{DFT Ref \cite{Chung}}  &substitution      & \multicolumn{2}{l}{work}
\\
\hline
J$_{NiNi}$	&2.99	&-13&77		&10&32  &9&99		\\
J$_{NiFe^O}$	&2.99	&-4&07	&-2&32  &-8&62	&-3&80	&J$_{YbNi}$	&-23&05	\\
J$_{NiFe^T}$	&3.50	&-24&80	&-23&61 &-23&61	&-23&60		\\
J$_{Fe^TFe^T}$	&3.60	&-8&22	&-32&25 &-32&32		\\
J$_{Fe^OFe^O}$	&2.99	&-19&12	&-11&62 &-19&37	&\multicolumn{2}{c}{ }	&J$_{YbFe^O}$	&-23&72	\\
J$_{Fe^TFe^O}$	&3.50	&-36&52	&-66&00 &-60&32	&-38&30	&J$_{YbFe^T}$	&-3&73
\\
\hline
\end{tabular}
\end{table}
\end{center}
\end{widetext}

\section{\label{sec:Summary}Summary and Conclusions}
In summary, a combined experimental and theoretical study was carried out through XPS, ND and magnetization measurements as well as spin-polarized DFT calculations to explain the electronic and magnetic structure of pure and Yb substituted strongly correlated oxide: NiFe$_{2-x}$Yb$_x$O$_4$. The  emphasis was on examining the strong correlation effect on the magnetic ordering of NFO and how the spin-half Yb impurity couples with the host to  affect the magnetization and T$_C$ of this inverse spinel ferrite.  
\vspace{6pt}
\\
The 3+ charge state of Yb (which substitutes the octahedral Fe) was confirmed from XPS measurements. The ND studies confirmed the ferrimagnetic ordering in NFO where spins at the octahedral sites (Ni$^{2+}$ and Fe$^{3+}$ cations) are aligned parallel while the spins at the tetrahedral sites (Fe$^{3+}$ cations) are antiparallel to those at the octahedral sites.  The magnetization measurements indicated a small decrease in the Curie temperature (T$_C$) from 853K to 836K with 7.5\% Yb substitution. From the estimation of exchange interaction strengths, it was found that the spins at the octahedral sites preferred to align antiparallel. However, stronger antiferromagnetic coupling with the spins at the tetrahedral sites forces them to parallel in order to avoid spin frustration.  By examining the effect of Hubbard U, it is concluded that this compound behaves as a mixed insulator under the Zaanen-Sawatzky-Allen classification scheme. Due to crystal field effect and strong correlation effect, the Yb-$f_{z(x^2-y^2)}$ orbital carries the unpaired spin which, unlike in the pure case, strengthens the antiferromagnetic coupling with the neighboring spins at the octahedral sites and significantly weakens the same with the spins at the tetrahedral site. Based on the estimation of magnetic coupling between the impurity and host spins, it was found that there was a net decrease in the antiferromagnetic interaction strength and as a consequence the T$_C$ of the compound is decreased.
\vspace{6pt}
\\
\section*{Acknowledgements}
The authors thank the HPCE  facility at IIT Madras for computations and University Grants Commission – Department of Atomic Energy Consortium for Scientific Research, Mumbai center, Bhabha Atomic Research Centre, Trombay, Mumbai-400 085 India for ND measurements. One of us, K. U., thanks Dr. Kamala Bharathi for the XPS measurements.

\bibliography{Paper}

\end{document}